\documentclass[times,twocolumn,final]{elsarticle}
\usepackage{medima}
\usepackage{framed,multirow}

\usepackage{amssymb}
\usepackage{latexsym}

\usepackage{amsmath}
\newcommand{\argmax}{\operatornamewithlimits{argmax}}

\usepackage[table,x11names]{xcolor}

\usepackage{url}
\usepackage{xcolor}

\usepackage{todonotes}

\usepackage{enumitem}

\usepackage{hyperref}

\usepackage{comment}
\definecolor{newcolor}{rgb}{.8,.349,.1}
\newcommand{\thickhline}{%
    \noalign {\ifnum 0=`}\fi \hrule height 1pt
    \futurelet \reserved@a \@xhline
}

\journal{Medical Image Analysis}

\begin{document}

\verso{Esther Puyol-Ant\'on \textit{et~al.}}

\begin{frontmatter}

\title{A Multimodal Deep Learning Model for Cardiac Resynchronisation Therapy Response Prediction}%

\author[1]{Esther \snm{Puyol-Ant\'on}\corref{cor1}}
\cortext[cor1]{Corresponding author: 
  E-mail address: esther.puyol\_anton@kcl.ac.uk (E. Puyol-Ant\'on)}
\author[1,2]{Baldeep S. \snm{Sidhu}}
\author[1,2]{Justin \snm{Gould}}
\author[1,2]{Bradley \snm{Porter}}
\author[1,
2]{Mark K. \snm{Elliott}}
\author[1,2]{Vishal \snm{Mehta}}
\author[1,2]{Christopher A. \snm{Rinaldi}}
\author[1]{Andrew P. \snm{King}}

\address[1]{School of Biomedical Engineering \& Imaging Sciences, King\textquotesingle s College London, UK}
\address[2]{Guy\textquotesingle{}s and St Thomas\textquotesingle{} Hospital, London, UK}

\received{XX}
\finalform{XX}
\accepted{XX}
\availableonline{XX}
%\communicated{S. Sarkar}

\begin{abstract}
We present a novel multimodal deep learning framework for cardiac resynchronisation therapy (CRT) response prediction from 2D echocardiography and cardiac magnetic resonance  (CMR) data. The proposed method first uses the `nnU-Net' segmentation model to extract segmentations of the heart over the full cardiac cycle from the two modalities. Next, a multimodal deep learning classifier is used for CRT response prediction, which combines the latent spaces of the segmentation models of the two modalities. At inference time, this framework can be used with 2D echocardiography data only, whilst taking advantage of the implicit relationship between CMR and echocardiography features learnt from the model. We evaluate our pipeline on a cohort of 50 CRT patients for whom paired echocardiography/CMR data were available, and results show that the proposed multimodal classifier results in a statistically significant improvement in accuracy compared to the baseline approach that uses only 2D echocardiography data. The combination of multimodal data enables CRT response to be predicted with 77.38\% accuracy (83.33\% sensitivity and 71.43\% specificity), which is comparable with the current state-of-the-art in machine learning-based CRT response prediction. Our work represents the first multimodal deep learning approach for CRT response prediction.

\end{abstract}

\begin{keyword}
%% Keywords
\KWD Multi-modality imaging \sep  Cardiac Resynchronisation Therapy \sep Treatment response prediction \sep Multi-view Deep learning 
\end{keyword}

\end{frontmatter}

%\linenumbers

%% main text
\section{Introduction}
\label{sec:Introduction}
Cardiac imaging techniques play a pivotal role in heart failure (HF) diagnosis, assessment of aetiology and treatment planning. Several modalities are available that are of relevance in patients with HF. Echocardiography is the first-choice imaging technique in the daily practice of cardiology as it is non-invasive, low-cost, easily available and provides most of the information required for the management and follow up of HF patients \citep{kirkpatrick2007echocardiography}. However, echocardiography has a number of limitations. First, image quality is heavily dependent on operator experience and expertise. Second, in normal clinical practice the images are two-dimensional (2D) meaning that geometrical assumptions are made in order to compute three-dimensional (3D) clinical metrics such as volumes and ejection fraction. Finally, echocardiography has limited spatial resolution with a relatively narrow field of view which can lead to poor endocardial definition.

Cardiac magnetic resonance (CMR) imaging is considered to be the gold standard in the evaluation of left ventricular (LV) function and is increasingly used in the assessment of HF due to its excellent temporal and spatial resolution and lack of ionizing radiation \citep{hundley2010accf}. However, CMR  is a time-consuming and expensive modality which requires significant technical expertise to operate, and hence has limited  availability in some geographical areas.

Recently, machine learning (ML), and more specifically deep learning (DL) techniques have shown promising performance in a range of medical image analysis tasks \citep{litjens2017survey}. In cardiac image analysis, most ML/DL techniques have considered the modalities of CMR and echocardiography in isolation. However, some works \citep{puyol2018regional,puyol2017multimodal,bruge2018multimodal} have shown that they can contain complementary information and so considering them in combination could have benefits in terms of performance. To date, these methods have been based upon traditional ML approaches such as multiview learning \citep{puyol2018regional}.

However, recently, inspired by the success of DL methods in other applications, 
multimodal deep learning (MMDL) \citep{ramachandram2017deep} has attracted significant research attention due to its ability to learn common features from multiple modalities, with the potential to exploit their natural strengths and reduce redundancies. In this paper we aim to use MMDL methods for prediction of response to Cardiac Resynchronisation Therapy (CRT), which is a common treatment for HF. Our aim is to use MMDL to produce an automated tool for CRT response prediction that utilises only echocardiography data as input at inference time whilst also exploiting multimodal (CMR and echocardiography) data at training time.

\subsection{Cardiac resynchronisation therapy}
\label{sbsec:CRT}
HF is a complex clinical syndrome associated with a significant morbidity and mortality burden. Cardiac remodeling is a pivotal process in the progression of HF, and it is defined as a change in the size, shape, or structure of one or more of the cardiac chambers. This remodelling can result in the development of dyssynchronous ventricular activation, which is often induced by electrical conduction delay in some regions of the LV and can lead to a decline in cardiac efficiency.

CRT is a common treatment for patients with heart failure with reduced ejection fraction (HFrEF) as it can restore LV electrical and mechanical synchrony. It has been shown to increase quality of life, improve functional status, reduce hospitalisation, improve LV systolic function and reduce mortality in properly selected patients \citep{bristow2004cardiac,cleland2005cardiac}. While CRT is an effective therapy, approximately 30-40\%  of patients treated with CRT gain little or no symptomatic benefit from the treatment \citep{yancy20172017, ponikowski20162016, mcalister2007cardiac, parsai2009toward}. The phenomenon of non-response to CRT is likely multi-factorial and related to patient selection criteria, CRT lead positioning and post-implant factors.
 
Current consensus guidelines \citep{authors20132013, ponikowski20162016} regarding selection for CRT focus on a limited set of patient characteristics including NYHA functional class, left ventricular ejection fraction (LVEF), QRS duration, type of bundle branch block, aetiology of cardiomyopathy and atrial rhythm (sinus, atrial fibrillation). The clinical research literature reveals a number of important insights into improving selection criteria, ranging from a lack of consensus regarding the definition of non-responders to technological limitations in the delivery of therapy. 

\cite{mullens2009insights} have previously described a post-implantation CRT optimisation clinic to investigate the causes of CRT non-response. They show that there were multiple common factors such as anemia, sub-optimal medical therapy, underlying narrow QRS duration and primary right ventricular dysfunction that could be identified pre-implantation and might help to improve outcomes and avoid implantation in unsuitable patients. Other factors that have been shown to be associated with increased response to CRT are strict left bundle branch block (LBBB) with type II contraction pattern \citep{jackson2014u} and presence of septal flash (SF)\footnote{An early inward motion of the ventricular septum.} and apical rocking \citep{stankovic2016relationship,marechaux2016role}. Other predictors of non-response to CRT that have been identified include ischemic cardiomyopathy, extensive scar, presence of right bundle branch block, absence of mechanical dyssynchrony, and poor LV lead placement (i.e. in a sub-optimal location) \citep{linde2012cardiac}. Despite more than 20 years of clinical development, a consensus definition of response and non-response to CRT has not been reached and it is necessary to better identify its causes for improving its results.

\subsection{Related work}
\label{sbsec:Related_work}
In this section, we provide an overview of the relevant literature on multimodal machine learning (Section \ref{sbsbsec:mmdl_related_work}) and the use of machine learning for CRT response prediction (Section \ref{sbsbsec:crt_related_work}).

\subsubsection{Multimodal machine learning}
\label{sbsbsec:mmdl_related_work}
Multimodal machine learning aims to build models that can process and relate information from multiple modalities. Compared to single modality machine learning techniques, learning from multimodal sources offers the possibility of capturing correspondences between modalities, reducing redundancies, and improving generalisation. Traditional multimodal machine learning approaches have included co-training algorithms \citep{brefeld2004co,muslea2000selective,yang2012information}, co-regularisation algorithms \citep{kan2015multi,sun2011multi}, margin consistency algorithms \citep{sun2013multi,chao2016consensus} and multiple kernel learning (MKL) \citep{gonen2011multiple}.

Several different DL algorithms have been proposed for multimodal learning. The  most common are: (1) variants of the deep Boltzmann machine, which have been proposed to model the joint distribution from different modalities' data \citep{srivastava2012multimodal,hu2013multimodal}; (2) extensions of classical autoencoders to discover correlations between hidden representations of two modalities \citep{wang2015deep,feng2014cross}; and (3) non-linear extension of the Canonical Correlation Analysis (CCA) algorithm using deep neural networks \citep{andrew2013deep}. However, in their originally proposed forms these models are not scalable to the number of features (pixels/voxels) typically present in medical images as they were based on fully-connected neural networks. Recently, some convolutional neural network (CNN)-based architectures that combine information from multiple sources for image and shape recognition have been proposed \citep{su2015multi,yao2017deepsense,wang20172d}.
With the aim of directly utilising the CMR and echocardiography data,
in this paper we employ a CNN-based architecture for deep multimodal classification.

\subsubsection{CRT response prediction}
\label{sbsbsec:crt_related_work}
A limited number of papers have investigated the use of ML to predict response to CRT. The literature can be mainly divided into approaches that use data from electronic health records (EHR) \citep{hu2019can,feeny2019machine,kalscheur2018machine,nejadeh2021predicting,ahmad2018machine, bernard2015impact}, approaches that use biomarkers derived from imaging data \citep{cikes2019machine,bernard2015impact,donal2019new,galli2021importance,chao2012intelligent,lei2019ventricular} and atlas-based approaches \citep{peressutti2017framework,duchateau2010atlas,sinclair2018myocardial}.

In the first category, the most common parameters used from the EHR are demographic information (e.g. sex, age, race), diagnosis codes (i.e. ICD9 and ICD10 codes), encounter information (i.e. visit type, length of stay), laboratory reports (e.g. lipids, glucose, creatinine), medication lists and cardiology reports (e.g. QRS duration, presence of LBBB, sinus rhythm). \cite{hu2019can} predicted CRT response in a cohort of 990 subjects using both structured and unstructured data from the EHR. The authors evaluated a variety of ML algorithms and showed that the gradient boosting classifier obtained the highest performance with a positive predictive value of 79\%. \cite{feeny2019machine} used a naive Bayes classifier with only 9 variables derived from the EHR and showed in a cohort of 455 subjects better CRT response prediction than current guidelines (area under the curve (AUC) = 0.7). Later, the same authors used Principal Components Analysis (PCA) followed by K-means clustering to predict CRT response using pre-and post- CRT 12-lead QRS waveforms \citep{feeny2020machine}. \cite{kalscheur2018machine} developed a random forest (RF) model for CRT patient survival prediction using 45 EHR features. Their model differentiated outcomes (AUC = 0.74) better than only current clinical discriminator features such as LBBB and QRS duration. \cite{lei2019ventricular} showed that the three features of QRS duration, LBBB, and non-ischemic cardiomyopathy achieved the highest accuracy (84.81\%) in identifying CRT responders when using a support vector machine (SVM) classifier. Finally,  \cite{ahmad2018machine} used a RF model to predict outcomes in 44,886 HF patients from the Swedish Heart Failure Registry. They also performed a cluster analysis to identify 4 distinct subgroups that differed significantly in outcomes and in response to therapeutics.

In the second category, the most common CMR and echocardiography-derived parameters used for CRT response prediction have been longitudinal, radial and circumferential strain \citep{cikes2019machine,bernard2015impact, donal2019new,chao2012intelligent} and right ventricular (RV) free wall strain and tricuspid annular plane systolic excursion (TAPSE) \citep{galli2021importance}. For example, \cite{cikes2019machine} used unsupervised MKL to combine EHR features with LV longitudinal strain derived from 2D echocardiography data to phenogroup patients with HF with respect to both outcomes and response to CRT. \cite{chao2012intelligent} computed radial peak strain from 2D echocardiography data on a cohort of 26 CRT patients, and used it with a SVM classifier to identify CRT responders with 95.4\% accuracy. \cite{galli2021importance} used a k-medoid algorithm with Gower distance to identify 16 features with good prediction of CRT response (AUC = 0.81) and outcomes (AUC = 0.84) in a cohort of 193 CRT patients. \cite{donal2019new} combined EHR data with 2D echocardiography-derived parameters in a RF algorithm for predicting response to CRT in a cohort of 54 HF patients. 

The works mentioned above focused on simple features derived from imaging data, and did not incorporate spatio-temporal information, which allows for a richer characterisation of cardiac function. Cardiac motion atlases have been previously used to exploit cardiac motion information from a cohort of subjects in a range of applications. Some works have used these atlases for CRT response prediction and to identify motion patterns that can be unique to CRT responders. \cite{duchateau2010atlas} built a cardiac motion atlas from 2D echocardiography data, and used it to detect septal flash and LV motion abnormalities in a cohort of CRT patients. \cite{peressutti2017framework} built a cardiac motion atlas from CMR imaging data and used supervised MKL to combine motion and non-motion features to predict CRT response, achieving approximately 90\% accuracy on a cohort of 34 patients. \cite{sinclair2018myocardial} built a cardiac motion atlas based on a novel approach to compute strain at different spatial scales in the LV from CMR imaging. A combination of PCA and linear discriminant analysis was used for identifying the spatial scales at which myocardial strain was most strongly predictive of CRT response. An accuracy of 86.7\% was achieved in identifying CRT responders in a cohort of 43 patients.

Until recently, DL had not been applied to predict CRT response. We have recently proposed such an approach in \cite{puyol2020interpretable}, which described a CMR-based pipeline based on a variational autoencoder (VAE) that allows CRT response prediction as well as the prediction of explanatory concepts to aid interpretability.

\subsection{Contributions}
\label{sbsec:Contributions}
In this paper we propose the first MMDL method for CRT response prediction. The method builds upon our recent work \citep{puyol2020interpretable} in which we proposed a DL framework for CRT response prediction based on CMR images, and extends it to exploit 2D CMR and 2D echocardiography data at training time. At inference time the CRT response prediction is made using only the echocardiography data whilst taking advantage of the implicit relationship between CMR and echocardiography features learnt from the model. This is the first ML model to learn features from multimodal imaging data for CRT response prediction. In addition, with the exception of our preliminary work \citep{puyol2020interpretable} it is the first DL model for CRT response prediction.

The remainder of this paper is organised as follows. In Section \ref{sec:Materials}, we describe details of the clinical data sets used for evaluation. In Section \ref{sec:Method} we describe the novel MMDL framework developed for CRT response prediction. In Section \ref{sec:Experiments} we present a thorough evaluation of the MMDL method, and Section  \ref{sec:Discussion} discusses the findings of this paper in the context of the literature and proposes potential improvements for future work.

\section{Materials}
\label{sec:Materials}
Four data sets were used for the training and validation of the MMDL model, and
these are described below:
\begin{enumerate}
    \item\textbf{UK Biobank (UKBB):} This database contains only CMR data and is used for pre-training the CMR segmentation model (see Section \ref{sbsc:seg}). In this work, we use a cohort of 700 healthy subjects, where the LV endocardial and epicardial borders and the RV endocardial border were manually traced at end diastole (ED) and end systole (ES) frames using the cvi42 software (version 5.1.1, Circle Cardiovascular Imaging Inc., Calgary, Alberta, Canada). CMR imaging was performed using a 1.5 T Siemens MAGNETOM Aera (see \cite{petersen2015uk} for further details of the image acquisition protocol).
    \item \textbf{EchoNet-Dynamic \citep{ouyang2020video}:} This database contains only echocardiography data and is used for pre-training the echocardiography segmentation model (see Section \ref{sbsc:seg}). Apical-4-chamber echocardiography images were acquired by skilled sonographers using iE33, Sonos, Acuson SC2000, Epiq 5G, or Epiq 7C ultrasound machines in a cohort of 10,030 patients.  For all subjects, the endocardial borders was manually traced at ED and ES frames. For further details of the image acquisition protocol see \cite{ouyang2020video}.
    \item  \textbf{Guy’s and St Thomas’ NHS Foundation Trust (GSTFT):} This database contains paired CMR and echocardiography data for a cohort of 50 HF patients and a cohort of 50 CRT patients. The GSTFT HF database was used to train and validate the CMR and echocardiography segmentation models (see Section \ref{sbsc:seg}). The GSTFT CRT database was used to train and validate the MMDL algorithm. Both studies were approved by the London Research Ethics Committee (11/LO/1232), all patients provided written informed consent for participation in this study and the research was conducted to the Helsinki Declaration guidelines on human research. CMR imaging was carried out on multiple scanners: Siemens Aera 1.5T, Siemens Biograph mMR 3T, Philips 1.5T Ingenia and Philips 1.5T and 3T Achieva. In this study, the cine CMR 4 chamber single-slice long-axis (la4Ch) data were used, which had a slice thickness between 6 and 10 mm and an in-plane resolution between 0.92x0.92mm\textsuperscript{2} and 2.4×2.4mm\textsuperscript{2}. For the GSTFT CRT cohort, 2D echocardiography imaging was acquired prior to CRT and at 6-months follow-up. For both cohorts, the ultrasound machines used were Philips IE33 and EPIQ 7C (Phillips Medical Systems, Andover, MA, USA) and the General Electric Vivid E9 (GE Health Medical, Horten, Norway), each equipped with a matrix array transducer. In this study, the apical 4Ch view was used for the development and evaluation of the MMDL model, and the apical 2Ch and 4Ch views were employed to estimate the left ventricular end-diastolic volume (EDV), end-systolic volume (ESV) and LVEF at 6-months follow-up for the GSTFT CRT cohort.  The echocardiography images had an in-plane resolution between 0.26x0.26mm\textsuperscript{2} and 0.62×0.62mm\textsuperscript{2}.
\end{enumerate}

\textbf{CRT volumetric response}: For the GSTFT CRT database, all patients fulfilled the conventional criteria for CRT (see Section \ref{sbsec:CRT}) and underwent CMR and 2D echocardiography imaging and clinical evaluation prior to CRT and at 6-months follow-up. All patients were classified as responders or non-responders based on volumetric measures derived from 2D echocardiography acquired at the 6-months follow-up evaluation \citep{authors20132013}. Patients were classified as responders if they had a reduction of $\geq$15\% in LV ESV after CRT, and were classed as non-responders otherwise. From the GSTFT CRT cohort, there were 32/50 patients who were classified as responders to CRT in this way and this information was used as the primary output label in training our proposed model.

\section{Methods}
\label{sec:Method}
\begin{figure}[!t]
\centering
\includegraphics[width=0.5\textwidth]{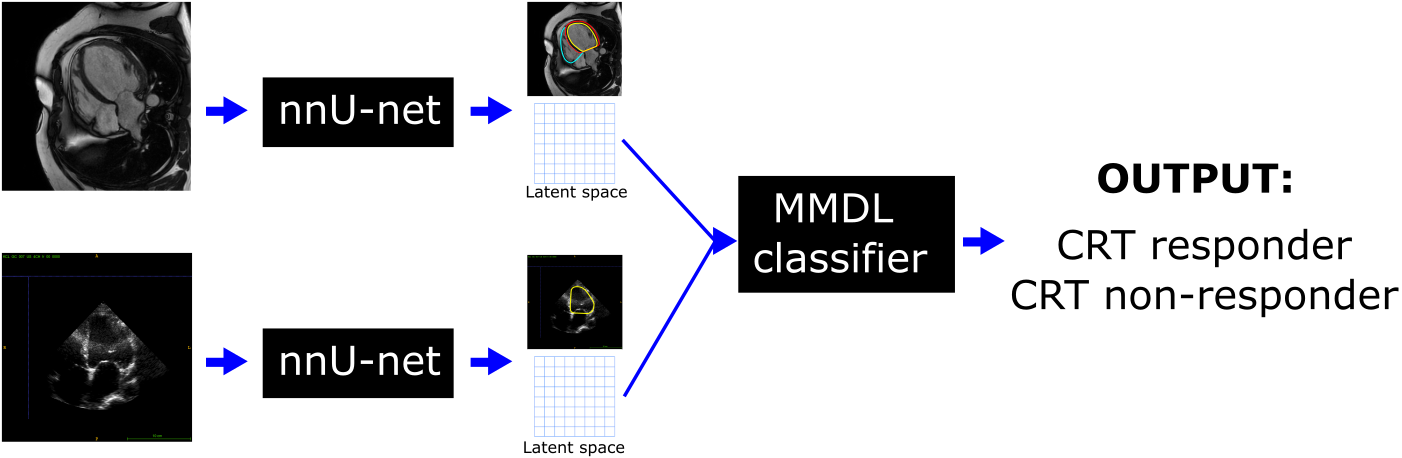}
\caption{Overview of the proposed framework to combine 2D CMR and echocardiography data for CRT response prediction. First, the `nnU-Net' architecture is used to extract segmentations of the heart over the full cardiac cycle from the two modalities. Next, a multimodal deep learning classifier is used for CRT response prediction, which combines the latent spaces of the `nnU-Net' models from the two modalities. At inference time, this framework can be used with 2D echocardiography data only, whilst taking advantage of the implicit relationship between CMR and echocardiography features learnt from the model.}
\label{fig:overview}
\end{figure}

An overview of the proposed MMDL framework is shown in Fig. \ref{fig:overview}. First, for each modality (i.e. CMR and echocardiography) a DL-based segmentation model is trained. Second, the latent spaces of these segmentation models are combined by the MMDL model and used for classification.   

In the following, Section \ref{sbsc:seg} briefly reviews the main steps involved in the automated segmentation of CMR and echocardiography images, Section \ref{sbsc:latent_space} describes how the latent spaces of the segmentation models are extracted and combined and Section \ref{sbsc:MMDL} introduces our MMDL classifier.

\subsection{Automatic segmentation network} 
\label{sbsc:seg}
\begin{figure}[!t]
\centering
\includegraphics[width=0.5\textwidth]{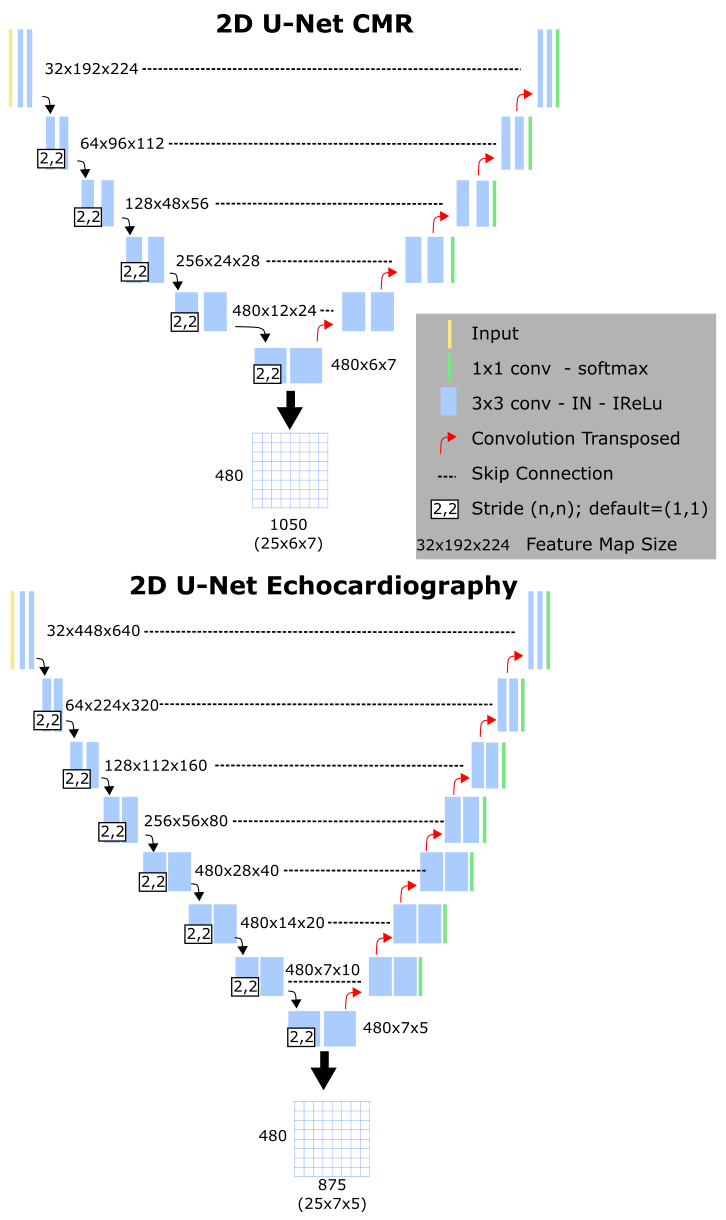}
\caption{The `nnU-Net' architecture. Top: the segmentation network used for CMR cine la4Ch images. Bottom: the segmentation network used for echocardiography apical 4Ch images. Below each figure, the generation of the latent space matrices is shown.}
\label{fig:seg}
\end{figure}

We used the `nnU-Net' architecture  \citep{isensee2021nnu} for automatic segmentation of CMR and echocardiography images in all frames through the cardiac cycle. In comparison to the standard U-net, the `nnU-Net' framework follows a holistic approach (topological parameters of the network architecture are automatically configured based on the training database) allowing for better generalisation. Figure \ref{fig:seg} shows an overview of the `nnU-Net' networks used for CMR and echocardiography data.\\

\textbf{CMR cine la4Ch segmentation:} The CMR segmentation model performed automated segmentation of the left ventricle blood pool (LVBP), left ventricular myocardium (LVMyo) and right ventricle blood pool (RVBP) from CMR la4Ch images. The network was trained with 1,400 images from the UK Biobank database (ED and ES frames) and 200 images (ED and ES frames) from the GSTFT HF cohort.\\

\textbf{Echocardiography apical 4Ch segmentation:} The echocardiography segmentation model performed automated segmentation of the left ventricle blood pool (LVBP) from echocardiography apical 4Ch images. The model was pre-trained using the EchoNet-Dynamic database \citep{ouyang2020video}, which includes 20,060 echocardiography images with annotations (ED and ES frames). To take into account the inter-vendor differences in intensity distributions, the segmentation model was then fine-tuned using 300 images (multiple time points from 50 echocardiography scans) from the GSTFT HF cohort.\\

\textbf{Implementation Details:} Both segmentation models were trained and evaluated using a five-fold cross-validation on the training set. As in \cite{isensee2021nnu}, the networks were trained for 1,000 epochs, where one epoch is defined as an iteration over 250 mini-batches. The batch sizes were 33 and 10 respectively for the CMR cine la4Ch and echocardiography apical 4Ch segmentation models. Stochastic gradient descent with Nesterov momentum ($\mu$=0.99) and an initial learning rate of 0.01 was used for learning network weights. The loss function used to train the `nnU-Net' model was the sum of cross-entropy and Dice loss. Data augmentation was performed on the fly and included techniques such as rotations, scaling, Gaussian noise, Gaussian blur, brightness, contrast, simulation of low resolution, gamma correction and mirroring. Please refer to \cite{isensee2021nnu} for more details of the network training.\\

\subsection{Generation of the latent space} 
\label{sbsc:latent_space}
The two segmentation models were used to segment the CMR and echocardiography images from the GSTFT CRT cohort in all frames through the cardiac cycle. To correct for variation in acquisition protocols between vendors, all images were first temporally resampled to  $T=25$ frames per cardiac cycle using piecewise linear warping based on cardiac timings \citep{puyol2018regional}. For each frame, the latent space of the segmentation network was stored and these were concatenated to generate a 2D matrix, in which rows correspond to the latent variables and columns to the different temporal frames - see Figure \ref{fig:seg} for an example case.

\subsection{Multimodal deep learning (MMDL)}
\label{sbsc:MMDL}
\begin{figure}[!t]
\centering
\includegraphics[width=0.5\textwidth]{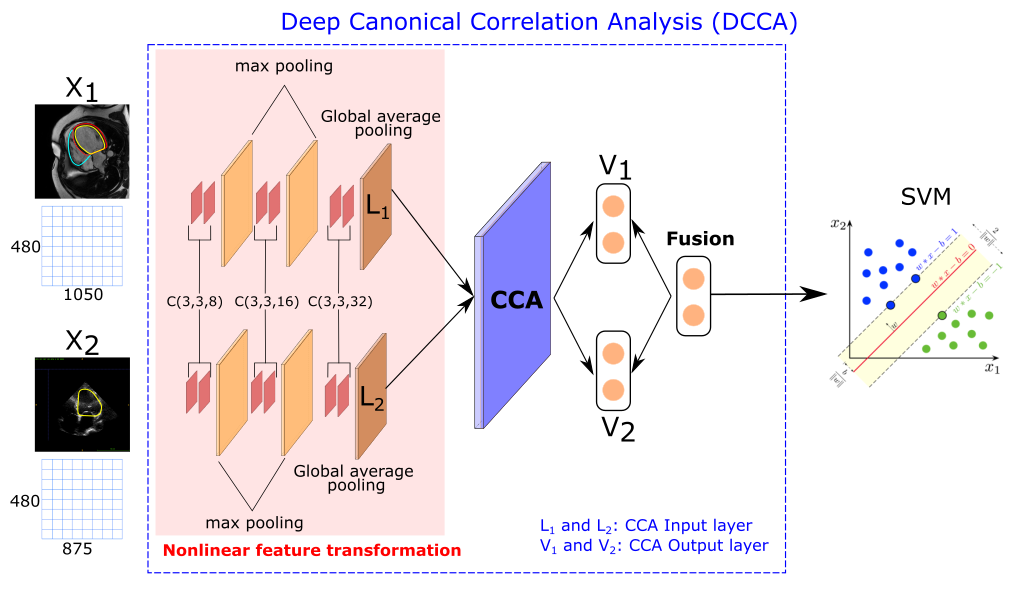}
\caption{Overview of the proposed MMDL method, which fuses the outputs of a DL model for each data view and applies a SVM classifier. C(m, n, k) denotes a convolutional layer with an m-by-n receptive field and k channels}
\label{fig:DCCA_SVM}
\end{figure}

We used the 2D Deep Canonical Correlation Analysis (DCCA) algorithm \citep{andrew2013deep,wang20172d} followed by a SVM classifier to develop our MMDL model for CRT response prediction. DCCA extends the linear CCA model by projecting multiple views of the data to a common latent space using a deep model with multiple branches, each corresponding to one view (see Fig. \ref{fig:DCCA_SVM}). Here we consider CMR and echocardiography images as two views of the same object, i.e. the heart. 

The 2D DCCA framework we employ is similar in concept to the configuration provided in \cite{wang20172d} and contains three parts:
\begin{enumerate}
    \item \textbf{Nonlinear feature transformation:} The first three layers are convolutional layers (i.e. CNNs): two 8-dims, two 16-dims and two 32-dims followed by a ReLU activation function. The size of all the receptive fields is 3×3 and the stride in all the layers is 1. A max pooling layer is inserted between each pair of sequences. The last layer is a global pooling layer, which enforces correspondences between feature maps and categories. The outputs of the last layer are defined as $L_1$ and $L_2$ in Fig. \ref{fig:DCCA_SVM}, and are the inputs of the CCA layer. 
    \item \textbf{CCA layer:} This layer  has the goal to jointly learn parameters for both views $X_1$ and $X_2$ to maximise corr($f_1(X_1)$,$f_2(X_2)$), where $f_1()$ and $f_2()$ are the nonlinear functions learnt by the networks from the previous step. We denote by $\theta_1$ the vector of all parameters from the nonlinear feature transformation of the first view, and similarly for $\theta_2$. These are determined according to the following objective:
    \begin{equation}
        (\theta_1^*, \theta_2^*) = \argmax_{(\theta_1, \theta_2)} \text{corr}(f_1(X_1;\theta_1),f_2(X_2;\theta_2)) 
    \end{equation}
    The parameters $\theta_1$ and $\theta_2$ of DCCA are trained to optimise this quantity using gradient-based optimisation. For the full derivation of the gradients we refer to the original DCCA paper by \cite{andrew2013deep}. The outputs of the CCA layer are $V_1$ and $V_2$.
    \item \textbf{Feature fusion} The fusion layer combines the outputs of the CCA layer ($V_1$ and $V_2$) as follows: $F_{fusion} = \alpha \cdot V_1 + \beta \cdot V_2$, where $\alpha$ and $\beta$  are the fusion weights. In our experiments, in order to balance the composition of features, we set   $\alpha=\beta=0.5$.
\end{enumerate}

The final part of the MMDL architecture is a binary SVM model with a radial basis function (RBF) kernel classifier, where the input is the output of the feature fusion layer and the output is a binary variable (i.e. CRT responders vs CRT non-responders). At inference time, we can use either the output of the CMR or echocardiography layer ($V_1$ or $V_2$) to simulate the scenarios in which only one of the modalities is available. \\

Compared to the standard DCCA framework, where the nonlinear feature transformations are extracted using fully connected layers, we have replaced these layers by  CNNs to take into account that our input data are 2D and to exploit the power of feature learning using CNNs.

\textbf{Implementation Details:} The segmentation models and the MDDL model were trained independently, to maximise the anatomical accuracy (and hence interpretability) of the segmentation models. The different parameters of the MMDL model were optimised using a grid search strategy. The 2D CCA model was trained for 500 epochs, with a batch-size of 10 images (generated from the latent space of the segmentation networks, see Section \ref{sbsc:latent_space}). We used the Adam optimiser with a momentum set to 0.9. For the 2D CCA model the parameters that were optimised were the size of the output layer k\textsubscript{DCCA}  (range $\in$ [5,10,15,20,25,30,35] and the learning rate, $lr$ (range $\in$ [0.01,0.001,0.0001]). The SVM hyperparameters that we optimised were $\gamma$ (range $\in$ [0.1,0.01,0.001,0.0001]) and the cost parameter $C$ (range $\in$ [1, 10, 100, 1000]). All  hyperparameters were optimised using a 5-fold nested cross-validation using the GSTFT CRT database (see Section \ref{sbsec:ev_metr} for more details).

\section{Experiments and results }
\label{sec:Experiments}
Two sets of experiments were performed. The first set of experiments (see Section \ref{sbsc:res_DSC}) aimed to validate the two segmentation models described in Section \ref{sbsc:seg}, while the second set of experiments (see Section \ref{sbsc:res_MMDL}) aimed to validate the proposed MMDL approach detailed in Section \ref{sbsc:MMDL}.

All experiments were carried out using the Python programming language with standard Python libraries Pytorch \citep{paszke2019pytorch} and scikit-learn.  Before describing the experiments in detail, we first describe the evaluation measures and comparative approaches used.

\subsection{Evaluation metrics and comparative approaches}
\label{sbsec:ev_metr}
For the segmentation networks, performance was evaluated using the Dice metric. For the MDDL model, a 5-fold nested cross-validation was used to validate its performance. In each nested fold, the predicted classes were stored and we then computed the overall classification balanced accuracy (i.e. the average of the accuracies obtained for each class individually), as well as the sensitivity (the proportion of CRT responders correctly classified) and the specificity (the proportion of CRT non-responders correctly classified). The balanced accuracy (BACC), sensitivity (SEN) and specificity (SPE) metrics are defined as:
\begin{itemize}[nosep]
\item Balanced accuracy (BACC): $\frac{1}{2}\left(\frac{TP}{TP+FN}+\frac{TN}{TN+FP}\right)$
\item Sensitivity (SEN): $\frac{TP}{TP+FN}$
\item Specificity (SPE): $\frac{TN}{TN+FP}$
\end{itemize}
where TP represents true positives, FP is false positives, FN is false negatives and TN is true negatives. \\

With the aim of evaluating the impact of using multimodal data for CRT response prediction, we compared the MMDL model with single modality classifiers. Due to the lack of prior work on image-based CRT response prediction, we chose to train and evaluate a number of different state-of-the-art CNN-based classifiers using the latent space matrices (see Fig. \ref{fig:seg}) of either the echocardiography or CMR segmentation models. We evaluated six different classifiers (AlexNet, DenseNet, MobileNet, ShuffleNet, SqueezeNet and VGG) for CRT response prediction using the two different modalities' latent space matrices. Each network was trained for 200 epochs with binary cross entropy loss, to classify between CRT responders and non-responders. During training, data augmentation was performed on-the-fly using random translations ($\pm$30 pixels), rotations ($\pm$90\textdegree), flips (50\% probability) and scalings (up to 20\%) to each mini-batch of images before feeding them to the network. The probability of augmentation for each of the parameters was 50\%. A nested cross validation was used to select the optimal learning rate, $lr$ (range $\in$ [0.01,0.001,0.0001]) and the optimal CNN classification network. We consider as a baseline the best such CNN trained using echocardiography data only, as it represents the current state-of-the-art in the use of echocardiography data alone for CRT response prediction.  Since CMR is considered to be the gold standard for analysis of cardiac function, we consider the best CNN technique trained using only CMR data as a reference technique.

In addition, we also compared the proposed approach with the DL-based baseline approach of our previous work \citep{puyol2020interpretable}. Using the GSTFT CRT cohort, we first tested the baseline VAE model implemented in \cite{puyol2020interpretable} using the short-axis (SAX) CMR data (as was the case in the original paper). Then, we trained the same model using the la4Ch CMR data. The first approach enabled a direct comparison with our previous framework, but using the GSTFT CRT cohort employed in this paper. Comparing the first and second approaches enabled us to assess the impact of using la4Ch data rather than SAX data for CRT response prediction. Finally, the second approach allowed a comparison of the technique described in \cite{puyol2020interpretable} to the MMDL model proposed in this paper (using la4Ch CMR data).

\subsection{Automated segmentation}
\label{sbsc:res_DSC}
For the CMR cine la4Ch segmentation model, the Dice metrics between automated and manual segmentations were 0.97 for the LV blood pool, 0.92 for the LV myocardium and 0.96 for the RV blood pool, which is in line with previous published methods \citep{leng2018computational,ruijsink2020fully}. For the echocardiography apical 4Ch segmentation network, similar to \cite{ouyang2020video}, the Dice metric between automated and manual segmentations was 0.96 for the LV blood pool.

\subsection{Evaluation of the multimodal deep learning framework} 
\label{sbsc:res_MMDL}
Table \ref{table:MMDL_results} shows the results of our experiments using the GSTFT CRT database.

As described in Section \ref{sbsec:ev_metr}, for the MMDL model a nested cross validation was used to ensure unbiased validation; the optimal parameters were: size of latent space k\textsubscript{DCCA}=25, learning rate lr=0.01, and the SVM parameters were $\gamma$=0.01 and $C$=10. The results of this method applied using CMR or echocardiography data are shown in the top row of Table \ref{table:MMDL_results} (i.e. \emph{MMDL}).

As a baseline approach, we report results for the VGG network trained and applied using only echocardiography data. VGG obtained the best BACC and SEN values compared to the other CNN classification networks. The optimal learning rate for this approach was $lr=0.001$. The results of this approach are shown in the second row of Table \ref{table:MMDL_results} (i.e. \emph{VGG\textsubscript{Echo}}).

As a reference approach (i.e. CNN trained and applied using only CMR data), VGG was again the best performing model, with  $lr=0.001$. These results are shown in the third row of Table \ref{table:MMDL_results} (i.e. \emph{VGG\textsubscript{CMR}}).

Finally, we compared to our previously published method \citep{puyol2020interpretable} which is trained and applied on CMR data only. In the original paper, the VAE model was trained using CMR SAX data. As discussed above, we evaluated this model using both SAX and la4Ch CMR data. These results are shown in the fourth and fifth rows of Table \ref{table:MMDL_results} (i.e. \emph{VAE\textsubscript{SAX}} and \emph{VAE\textsubscript{LAX}}).

Student’s t-tests (99\% confidence) were used to compare the performance of the baseline approach with the other approaches.

\begin{table}[ht] 
\caption{Balanced Accuracy (BACC), Sensitivity (SEN) and Specificity (SPE) of the proposed and comparative methods and Student’s t-test (99\% confidence) results. The CMR and Echocardiography column headings indicate which data were used to apply the model. The first group corresponds to the results of the MMDL method for CRT response prediction, the second group corresponds to the results of the baseline method, and the third group corresponds to the  comparative methods. An asterisk indicates a statistically significant improvement in accuracy over the baseline comparative approach. Bold-italic text indicates the method with the overall best classification accuracy, bold indicates the indicates the method with the highest classification accuracy for the la4Ch data, and dash indicates that this approach cannot be applied for this modality.}
\centering
\resizebox{0.5\textwidth}{!}{
\begin{tabular}{l ccc | ccc}
& \multicolumn{3}{c}{\textbf{CMR}} & \multicolumn{3}{|c}{\textbf{Echocardiography}} \\  \cline{2-7}
 & \textbf{BACC(\%))} & \textbf{SEN(\%)} & \textbf{SPE(\%)} & \textbf{BACC(\%)} &\textbf{SEN(\%)} & \textbf{SPE(\%)}\\ \hline
\multicolumn{7}{c}{\textit{1. Multimodal deep learning approach}} \\
MMDL  & \textbf{81.19*} & 86.21 & 76.19 & \textbf{77.38*} & 83.33 & 71.43\\ \hline \hline
\multicolumn{7}{c}{\textit{2. Baseline approach (echocardiography-trained)}} \\
VGG\textsubscript{Echo} & - & - & - & 70.26 & 75.00 & 65.52\\ \hline
\multicolumn{7}{c}{\textit{3. Single modality (CMR-trained) approaches}} \\
VGG\textsubscript{CMR} & 71.98\phantom{*} & 68.97 & 75.00 & - & - & - \\
VAE\textsubscript{SAX} & \textbf{\textit{82.64*}} & 87.50  &  77.78  & - & - & - \\
VAE\textsubscript{la4Ch} & 78.30* & 84.38  &  72.22 & - & - & - \\
\end{tabular}}
\label{table:MMDL_results}
\end{table}

The results show that the MMDL framework is capable of performing CRT response prediction using data from either view with similar accuracy, sensitivity and specificity. We assessed the statistical significance of the BACC using the proposed method compared to the performance of the baseline approach of VGG\textsubscript{Echo} trained on echocardiography data and we show that our proposed method outperforms the baseline approach. For the comparative approaches, we can see that the multimodal classifier algorithm has a higher accuracy compared to the single modality methods except for the VAE-based approach trained with SAX data (BACC 82.64\% compared to 81.19\%). 

For the VAE-based approach, if we compare the models trained using SAX and la4Ch data, we can see that there is a slight decrease in accuracy when using la4Ch data. As mentioned in Section \ref{sbsec:ev_metr}, the VAE\textsubscript{SAX} results are only reported for comparison to our previous work \citep{puyol2020interpretable}. However, this method cannot be directly compared to the proposed MMDL method as it is trained using different data, and paired echocardiography data would not be available for the SAX view. We hypothesize that the difference can be explained by the different regions covered by the two modalities: the SAX data focuses on the basal region of the heart where there is more motion, whilst the la4Ch images focus on a cross section of the heart which may not capture some motion that can be important for CRT response prediction.

From Table \ref{table:MMDL_results} we can also see that using only la4Ch CMR data we achieve the highest accuracy with the MMDL model. Based on these results, we conclude that the highest accuracy is achieved using our proposed multimodal deep learning framework, whether using only echocardiography data or CMR data as input. We also conclude that including the CMR data into the training (as well as the echocardiography data) improves the performance of the MMDL classifier.\\

\section{Discussion and conclusion}
\label{sec:Discussion}
We have proposed a novel MMDL method based on DCCA that has the ability to predict CRT response using only echocardiography data but at the same time taking advantage of the implicit relationship between CMR and echocardiography. To the best of our knowledge, this is the first time that multimodal imaging data has been used directly for CRT response prediction. Previous works have mainly focused on imaging-derived parameters or atlas-based approaches, where manual input is required.
Compared to our previous work \citep{puyol2020interpretable}, the proposed approach allows CRT response to be predicted using only the widely available and cheap modality of echocardiography. Some previous work has proposed a multimodal ML model that exploits echocardiography alone at inference time \citep{puyol2018regional}. Compared to this work, the proposed approach makes direct use of 2D CMR and 2D echocardiography imaging data, rather than the results of motion tracking algorithms, which reduces the complexity of building the spatio-temporal model and eliminates the effect of potential errors in the motion tracking. In addition, the proposed framework is fully automated while the framework proposed in \cite{puyol2018regional} required manual delineations of the CMR and echocardiography data.

Our results showed that the use of the MMDL algorithm resulted in a statistically significant increase in classification accuracy compared to the use of only echocardiography or CMR data, and that our technique for prediction of CRT response (i.e. 83.33\%/71.43\% sensitivity/specificity) is comparable with the current state-of-the-art. For example, \cite{peressutti2017framework} reported 100\% sensitivity and 62.5\% specificity for predicting CRT responders based on  a combination of a cardiac motion atlas with non-motion data obtained from several sources. \cite{sohal2014prospective} reported 85\% sensitivity and 82\% specificity for predicting CRT responders based on volume-change systolic dyssynchrony index. Importantly, both of these approaches required significant manual intervention.

It is important to consider the clinical interpretation of sensitivity and specificity in the context of CRT response. Sensitivity refers to the proportion of CRT responders correctly identified and specificity refers to the proportion of CRT non-responders correctly identified. In our application, we are interested in achieving the highest possible sensitivity to ensure that there are no subjects that are denied a treatment that would have resulted in symptomatic benefit. Our technique achieved a sensitivity of 83.33\%, meaning that only $\sim$17\% of such cases
would have been missed, and specificity was 71.43\%, meaning that almost three quarters of non-responders would be spared the unnecessary treatment, which is not without risk. To foster clinical translation and promote clinical trust, it is important that sensitivity is as high as possible, preferably 100\%, and this will be the focus of future work.

Increasing the response rate to CRT has been the focus of many studies and the results to date have been conflicting. This is because CRT treatment planning is multi-factorial and related to patient selection criteria, CRT lead positioning and post-implant factors. Future work will aim to incorporate other clinical parameters such as pacing leads, presence of scar and septal flash to have a more
complete and clinically useful pipeline.\\

Another area for future work is to validate the approach on a larger multi-centre database of CRT patients, to better evaluate the accuracy and robustness of the proposed framework. Furthermore, we would like to test the same pipeline for predicting response to other types of treatment, such as pharmacology, and try to incorporate uncertainty techniques to be able to estimate confidences in CRT response prediction. This would move us closer to our vision of an interpretable tool that can be used to support the decision making of cardiologists when selecting HF patients for different types of treatment.

\section*{Acknowledgments}
This work was supported by the EPSRC (EP/R005516/1, EP/P001009/1), by core funding from the Wellcome/EPSRC Centre for Medical Engineering (WT203148/Z/16/Z), and by the NIHR Cardiovascular MedTech Co-operative award to the Guy’s and St Thomas’ NHS Foundation Trust. EPA and APK acknowledge financial support from the Department of Health through the National Institute for Health Research (NIHR) comprehensive Biomedical Research Centre award to Guy’s \& St Thomas’ NHS Foundation Trust in partnership with King’s College London. The views expressed are those of the author(s) and not necessarily those of the NHS, the NIHR, EPSRC, or the Department of Health. This research has been conducted using the UK Biobank Resource (application number 17806) on a GPU generously donated by NVIDIA Corporation. The UK Biobank data are available for approved projects from https://www.ukbiobank.ac.uk/.

%%Harvard
\bibliographystyle{model2-names.bst}\biboptions{authoryear}
\bibliography{refs}

%\section*{Supplementary Material}

\end{document}